\documentclass[reprint, aps,prd,twocolumn,showpacs,showkeys, 10pt, longbibliography, nolinenumbers, floatfix]{revtex4-2}

\usepackage{amsmath}
\usepackage{graphicx}
\usepackage{float}
\usepackage{dcolumn}
\usepackage{multirow}
\usepackage{natbib}
\usepackage{changes}
\usepackage[colorlinks = true,linkcolor = blue,urlcolor  = blue,citecolor = blue,anchorcolor = blue]{hyperref}
\usepackage{enumerate}
\usepackage{verbatim}
\usepackage{orcidlink}
\usepackage{microtype}
\usepackage{times}

\begin{document}

\title{Simultaneous explanation of XTE J1814-338 and HESS J1731-347 objects \\ using ${K^{-}}$ and ${\bar{K}^{0}}$ condensates}

\author{M. Veselsk\'y$^1$\orcidlink{0000-0002-7803-0109}}
\email{Martin.Veselsky@cvut.cz}
\author{V. Petousis$^1$\orcidlink{0000-0002-5575-6476}}
\email{vlasios.petousis@cvut.cz}
\author{P.S. Koliogiannis$^{2,3}$\orcidlink{0000-0001-9326-7481}}
\email{pkoliogi@phy.hr}
\author{Ch.C. Moustakidis$^3$\orcidlink{0000-0003-3380-5131}}
\email{moustaki@auth.gr}
\author{J. Leja$^4$\orcidlink{0000-0003-1798-116X}}
\email{jozef.leja@stuba.sk}

\affiliation
{$^1$Institute of Experimental and Applied Physics, Czech Technical University, Prague, 110 00, Czechia \\
$^2$ Department of Physics, Faculty of Science, University of Zagreb, Bijeni\v cka cesta 32, Zagreb, 10000, Croatia \\
$^3$Department of Theoretical Physics, Aristotle University of Thessaloniki, 54124, Thessaloniki, Greece \\
$^4$Faculty of Mechanical Engineering - Slovak University of Technology in Bratislava, Bratislava, 81231, Slovakia 
}

\begin{abstract}
The recent observation of the compact star XTE J1814-338 with a mass of $M=1.2^{+0.05}_{-0.05}~{\rm M_{\odot}}$ and a radius of $R=7^{+0.4}_{-0.4}$ km, together with the HESS J1731-347, which has a mass of $M=0.77^{+0.20}_{-0.17}~{\rm M_{\odot}}$ and a radius of $R=10.4^{+0.86}_{-0.78}$ km, shows they provide evidence for the possible presence of exotic matter in the core of neutron stars and significantly enhance our understanding of the equation of state for the dense nuclear matter. In the present study, we investigate the possible existence of negative charged kaons and neutral anti-kaons in neutron stars by employing the relativistic mean field model with first order kaonic (${K^{-}}$ and ${\bar{K}^{0}}$) condensates. To the best of our knowledge, this represents a first alternative attempt aimed to explain the bulk properties of the XTE J1814-338 object and at the same time the HESS J1731-347 object, using a mixture of kaons condensation in dense nuclear matter. In addition, we compare our analysis approach with the recent observation of PSR J0437-4715 and PSR J1231-1411 pulsars, proposing that to simultaneously explain the current variety of astrophysical objects, it is essential to resurrect a scenario of two distinct branches, each corresponding to a different composition of nuclear matter.
\end{abstract}

\maketitle
\noindent {\it{Introduction.}} We know today that neutron stars (NSs) are compact celestial objects with densities approaching and exceeding the nuclear saturation density, typical for atomic nuclei. Bulk properties of NSs, such as the maximum mass that exceeds $2M_{\odot}$ and the radius at canonical mass around $1.4M_{\odot}$ ($R = 12-13~{\rm km}$), can be reproduced by a nuclear equation of state (EoS) considering neutrons, protons and electrons, even if further degrees of freedom are introduced, such as muons, hyperons and kaons~\cite{Glendenning-2000,Haensel-2007}. In particular, kaon condensation in dense matter was suggested by Kaplan and Nelson~\cite{Kaplan-1986}, and has been discussed in many other publications~\cite{BR-1994,Pandharipande-1995,Wass-1996, Wass-1997}. Because of the attraction between $K^{-}$ and nucleons, its energy decreases with increasing density, and eventually, if it drops below the electron chemical potential in a ${\beta}$-equilibrium NS matter, a Bose condensate of $K^{-}$ will appear. The kaon-nucleon interaction in vacuo has been described by Brown \textit{et al.}~\cite{BR-1994} using an effective Lagrangian based on chiral perturbation theory. 

The $K^{-}N$ interaction is well described and fortunately is not affected much by resonances as the $K^{-}p$ interaction. Kaiser \textit{et al.}~\cite{Wass-1996, Wass-1997} have used energy dependent $K^{-}N$ amplitudes calculated in a coupled-channel scheme starting from the chiral SU(3) effective Lagrangian. They correct for correlation effects in a nuclear medium and find that $K^{-}$ condenses at densities higher than $4\rho_{0}$, where $\rho_{0}=0.16~{\rm fm^{-3}}$ is the normal density of nuclear matter. This should be compared to the central density of $4\rho_{0}$ for a NS of mass $1.4M_{\odot}$ according to the estimates of Wiringa \textit{et al.}~\cite{Fiks-1988} using realistic models of nuclear forces. The condensate could change the structure and affect the maximum masses and cooling rates of massive NSs.

Within the extremely dense core of NSs, the density and pressure can reach quite high values, energetically allowing the formation of kaons in order to balance the Fermi energy and reduce the overall energy density. As the density in NSs increases, electrons become highly degenerate, and beyond a certain threshold, it becomes energetically favorable for $K^{-}$ to replace electrons as charge neutralizing agents. This procedure introduces an excess of strangeness, as $K^{-}$ contains an anti-strange quark ($\bar{s}$). With further increase in density, the energy conditions favorable the formation of $\bar{K}^{0}$ ($m_{\bar{K}^{0}}c^{2} > m_{ K^{-}}c^{2}$), which, due to their lack of electric charge, introduce additional strangeness without altering the charge neutrality. The presence of $\bar{K}^{0}$ leads to further softening of the EoS at high energy density enabling the formation of strangeness-rich compact object. 

Until very recently, NSs or compact objects with mass below the canonical mass of $1.4M_{\odot}$ were unclassified. Specifically, within the last year the compact object HESS J1731-347 was reported with a mass of $M=0.77^{+0.20}_{-0.17}~{\rm M_{\odot}}$ and a radius of $R=10.4^{+0.86}_{-0.78}$ km~\cite{HESS-2023}, while a new analyses of the compact object XTE J1814-338~\cite{Kini-2024, Baglio-2013} revealed a mass of $M=1.2^{+0.05}_{-0.05}~{\rm M_{\odot}}$ and a radius of $R=7^{+0.4}_{-0.4}$ km. In view of the above, several light compact objects were reported which can hardly be explained simultaneously and at the same time reproduce the NS maximum mass constraint. Furthermore, besides the HESS J1731-347 and XTE J1814-338 objects, also a new analysis of PSR J1231-1411~\cite{Salmi-2024} ($M=1.04^{+0.05}_{-0.03}~{\rm M_{\odot}}$ and radius $R=12.6^{+0.3}_{-0.3}$ km) appeared recently, noting that the authors report problems related to the convergence in the fitting procedure and the uncertainty might be larger than reported. Such variety of light compact objects obviously requires variety also in physics scenarios, characterized by different structure. 

Several studies attempt to reproduce the properties of the XTE J1814-338 using a concept of hybrid star. Specifically, Pitz and Schaffner-Bielich \cite{Pitz-2024} studied the possibility of the XTE J1814-338 being a bosonic star with a nuclear matter core, while Yang \textit{et al.}~\cite{Yang-2024} investigated the scenario of a strange star admixed with mirror dark matter. Recently, Lopes and Issifu~\cite{Lopes-2024} claimed that  this object can be explained also as a dark matter admixed neutron star. Notably, the idea of ultra-compact stars has been studied by Li \textit{et al.}~\cite{Sedrakian-2023} before the results of Ref.~\cite{HESS-2023}. 
Recent work of Laskos-Patkos and Moustakidis~\cite{Laskos-2024} achieves the agreement with the XTE J1814-338 by applying an assumption of first order phase transition with specifically chosen transition density and energy density jump. 
Another possibility is the existence of strangeness, which can be considered
as an additional degree of freedom, the amount of which might be determined by the formation scenario of a specific object. In general, as in nuclear reactions, production of strangeness requires higher densities and energies (temperatures). Thus, compact objects with strangeness can be produced in collapse of more massive stellar objects, where higher temperatures and densities allow copious production of strangeness~\cite{Haensel_2007, Brown-Rho_2008}. 

Therefore, we introduce an EoS suitable to describe both the bulk properties of NSs and the ones of ultra-light compact objects, by considering two distinct branches: (a) a nucleonic branch that allows the description of the current astrophysical constraints on NSs, and (b) an exotic branch consisting of a mixture of kaons, that properly fascilitate the characteristics of ultra-light compact objects{~\cite{Haensel_2007}. We note that in our previous work~\cite{Veselsky-2024}, an effort was made to reproduce the properties of the HESS J1731-347 object using only $K^{-}$ bosonic condensate. In this study, we extend this argument by incorporating also a $\bar{K}^{0}$ condensate along with the $K^{-}$ one.

\vspace{0.2cm} 
\noindent{\it{Nuclear theoretical framework.}}
In the relativistic mean field (RMF) theory, using the extended Dirac-Hartree approximation, the energy density and pressure of neutron matter is given by the following expressions \cite{Serot-1997}:
\begin{align}
{\cal E}_{N}&=\frac{(\hbar c)^3g_{\omega N}^2}{2(m_{\omega}c^2)^2}n_N^2+ \frac{(\hbar c)^3(\frac{g_{\rho N}}{2}){^2}}{2(m_\rho c^2)^2}\rho_I^2 \nonumber\\
 &+ \frac{(m_{\sigma}c^2)^2}{2g_{\sigma N}^2(\hbar c)^3}(M_Nc^2-M_N^*c^2)^2\nonumber\\
 &+ \frac{\kappa}{6g_{\sigma N}^3}(M_Nc^2-M_N^*c^2)^3+\frac{\lambda}{24g_{\sigma N}^4}(M_Nc^2-M_N^*c^2)^4\nonumber\\
 &+\sum_{i=n,p} \frac{\gamma}{(2\pi)^3}\int_0^{k_Fi} 4\pi k^2 \sqrt{(\hbar c k)^2+(m_i^* c^2)^2}dk,
 \label{RMF-E-1}
\end{align}
\begin{align}
{\cal P}_{N}&=\frac{(\hbar c)^3g_{\omega N}^2}{2(m_{\omega}c^2)^2}n_N^2+ \frac{(\hbar c)^3 (\frac{g_{\rho N}}{2}){^2}} {2(m_\rho c^2)^2}\rho_I^2 \nonumber \\
 &-\frac{(m_{\sigma}c^2)^2}{2g_{\sigma N}^2(\hbar c)^3}(M_Nc^2-M_N^*c^2)^2\nonumber\\
 &+ \frac{\kappa}{6g_{\sigma N}^3}(M_Nc^2-M_N^*c^2)^3+\frac{\lambda}{24g_{\sigma N}^4}(M_Nc^2-M_N^*c^2)^4\nonumber\\
 &+\sum_{i=n,p}  \frac{1}{3}\frac{\gamma}{(2\pi)^3}\int_0^{k_Fi} \frac{4\pi k^2}{\sqrt{(\hbar c k)^2+(m_i^* c^2)^2}}dk,
 \label{RMF-P-1}
\end{align}
where ${\cal E}_{N}$ is the energy density, ${\cal P}_{N}$ is the pressure, $g_{\sigma N}$, $g_{\omega N}$ and $g_{\rho N}$ are the couplings of the scalar boson, vector boson, and iso-vector $\rho$-meson to nucleons, respectively, $m_{\sigma}$, $m_{\omega}$ and $m_\rho$ are the rest masses of scalar and vector bosons, and $\rho$-meson, respectively, the term $\rho_I$ involves the difference between the proton and neutron densities (important for finite nuclei), $\kappa$ and $\lambda$ are the couplings of the cubic and quartic self-interaction of the scalar boson, $M_N$ and $M_N^*$ are the rest mass and the effective mass of the nucleon,
$n_N$ is the nucleonic density, $k_F$ is the Fermi momentum of nucleons at zero temperature and $\gamma$ is the degeneracy, with value $\gamma= 4$ for symmetric nuclear matter and $\gamma= 2$ for neutron matter (used in this investigation). 

The kaon condensate was introduced according to the first order kaon condensate (FOKC) model of Glendenning and Schaffner-Bielich~\cite{Glendenning-1998, Glendenning-1999}. It specifically considers $K^{-}$ particles which can play similar role as electrons in the NS matter. The kaon potential was fixed by the value at saturation density $\rho_0$ of symmetric nuclear matter~\cite{Knorren-1995, Hong-2024}:
\begin{equation}
U_{K}(\rho_0) = - g_{\sigma K} \frac{(M_{N}-M_{N}^{*}(\rho_0)) c^{2}}{g_{\sigma N}}-(\hbar c)^3 g_{\omega K} g_{\omega N} \frac{\rho_0}{m_{\omega}^{2} c^{4}},
\label{uk0}
\end{equation}
where $g_{\sigma K}$ and $g_{\omega K}$ are couplings of $\sigma$ and $\omega$ mesons to $K^{-}$. The $K^{-}$ chemical potential at a given baryonic density was evaluated as: 
\begin{eqnarray}
\mu_{K-}(\rho,x_{p}) &=& m_{K} - g_{\sigma K} \frac{(M_{N}-M_{N}^{*}(\rho_{N},x_{p})) c^{2}}{g_{\sigma N}} \nonumber\\  
&-& (\hbar c)^3 g_{\omega K} g_{\omega N} \frac{\rho_{N}}{m_{\omega}^{2} c^{4}} \nonumber\\  
&-& (\hbar c)^3 g_{\rho K} g_{\rho N} \frac{\rho_N}{m_{\rho}^{2} c^{4}}(1 - 2 x_{p}),
\label{muk}
\end{eqnarray}
where $g_{\rho K}$ is a coupling of $\rho$ meson to $K^{-}$ and $x_{p}$ is the proton fraction. The formula for the chemical potential of $\bar{K}^{0}$ is similar to the one of $K^{-}$, where the only difference is the opposite sign of the last term. The effective mass $M_{N}^{*}(\rho,x_{p})$ is approximated by parabolic dependence on $x_{p}$ between values for symmetric nuclear matter and pure neutron matter. The value of $\mu_{K}(\rho,x_{p})$ is then used to calculate the conditions for the chemical equilibrium of the system: 
\begin{eqnarray}
  \mu_{n} - \mu_{p}  = \mu_{e} = \mu_{K}, \nonumber \\
  \rho_{p}  = \rho_{e} + \rho_{K},
\label{betaeq}
\end{eqnarray}
which also provides the electron and kaon densities. When the kaons chemical potential drops to zero, $\bar{K}^{0}$ particles also produced in dense nuclear matter. Finally, the energy density for kaons can be expressed as: 
\begin{equation}
\epsilon_{K} = \mu_{K} \rho_{K}.
\label{ek}
\end{equation}

Similar to Refs.~\cite{Glendenning-1998, Glendenning-1999}, both kaon condensates are not considered to contribute to pressure.

\begin{figure}[t]
\centering
\includegraphics[width=9.cm,height=8.cm]{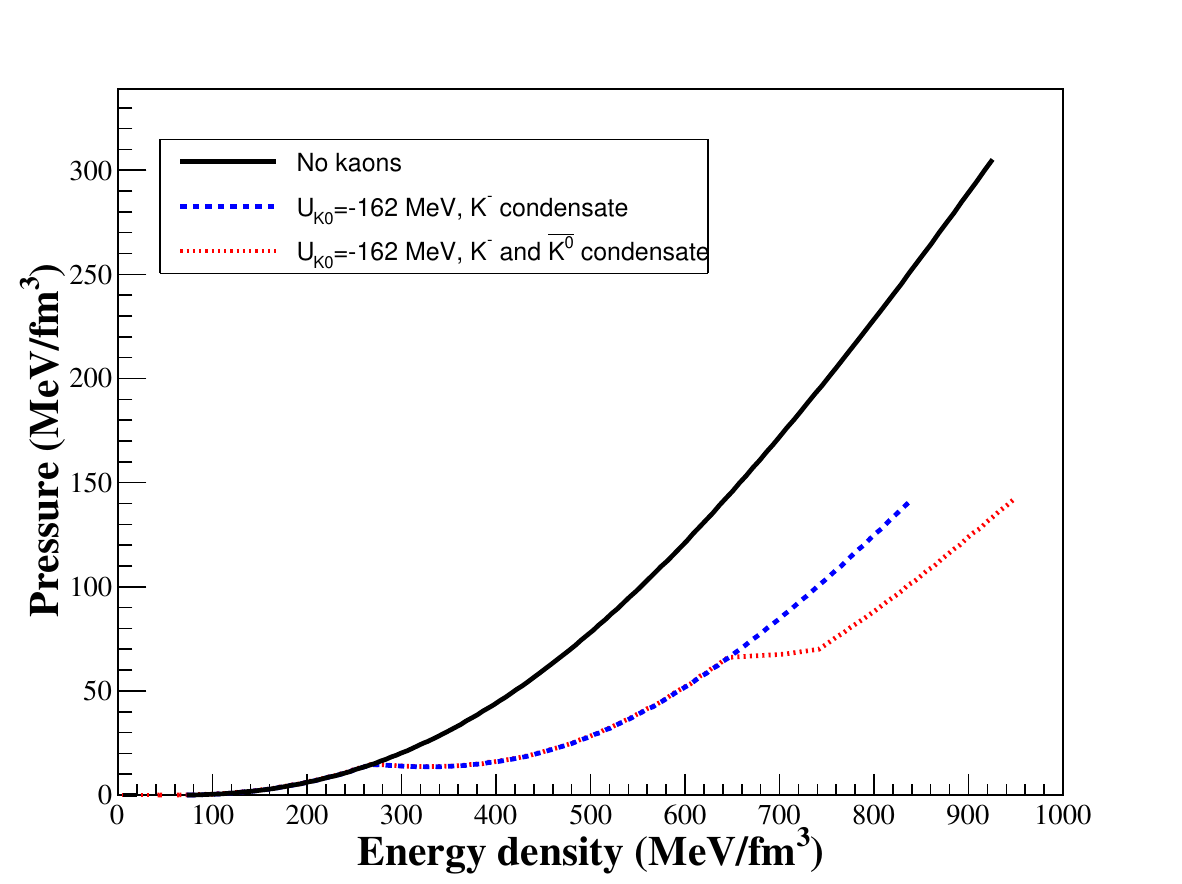}
\caption{Equation of state for no kaons (black solid line) and for the kaonic potential $U_{K0}$= - 162 MeV using $K^{-}$ (blue dashed line) and both the $K^{-}$ and $\bar{K}^{0}$ condensates (red dotted line).}
\label{fgeos}
\end{figure}

\vspace{0.2cm} 
\noindent{\it{Results and discussion.}} In the present work we extend the previous investigation of Ref.~\cite{Veselsky-2024} by introducing a ${\bar{K}^{0}}$ condensate at higher densities than the emerge of $K^{-}$, in order to soften the EoS and simultaneously describe the properties of the XTE J1814-388 and HESS J1731-347 light compact objects. This might mean that the Mass-Radius (M-R) diagram of the compact objects can have a "kaonic" branch, deviating from the M-R dependence of standard nucleonic NS. In particular, we use the RMF model~\cite{Serot-1997} and a set of parameters suitable to describe the nucleonic matter, which reproduces existing constraints on maximum mass of NS and its radius at canonical mass $1.4 M_{\odot}$. As in our previous work~\cite{Veselsky-2024}, the ${K^{-}}$ condensate implemented according to Glendenning and Schafner-Bielich~\cite{Glendenning-1998, Glendenning-1999}, and furhter extended by the introduction of ${\bar{K}^{0}}$ condensate. Specifically, the ${\bar{K}^{0}}$ condensate can emerge when its chemical potential ($\omega_{\bar{K}^{0}}$ = $\omega_{K^{-}}$ + 2 $g_{\rho K} R_{03}$) reaches zero~\cite{Glendenning-1998, Glendenning-1999}. It worth mentioning that since the considered kaons form an isospin doublet, the model leads to an equal number of densities for both condensates. By employing the above parametrization, we calculated the bulk properties of compact stars for various values of the kaonic potential. The conditions for $\beta$-equilibrium were assumed in standard way as in Eq.~\eqref{betaeq}. 

\begin{figure}[t]
\centering
\includegraphics[width=9.cm,height=8.cm]{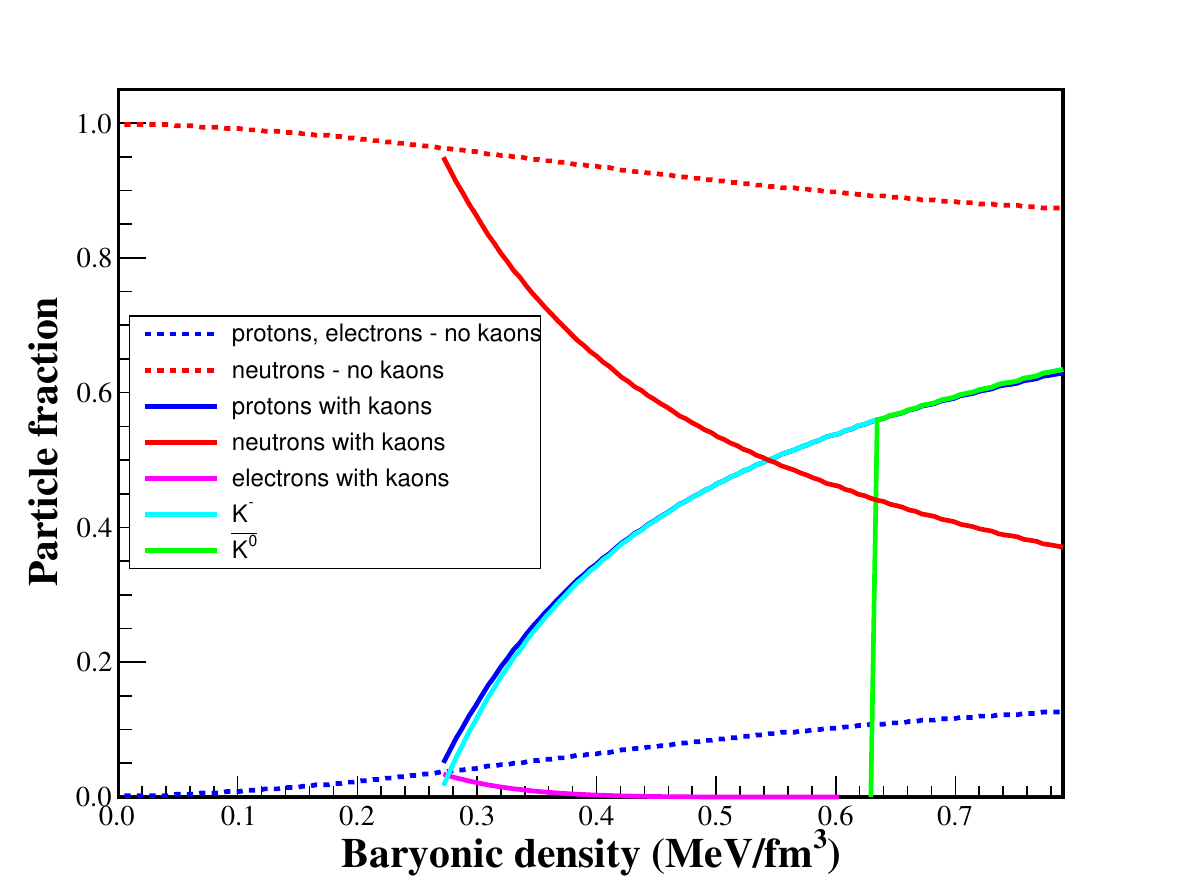}
\caption{Particle fractions as a function of the baryonic density for the RMF - EoS with kaon potential U$_{K0}$= -162 MeV.}
\label{fig:pf_rmf}
\end{figure}

\begin{figure}
\centering
\includegraphics[width=9.cm,height=8.cm]{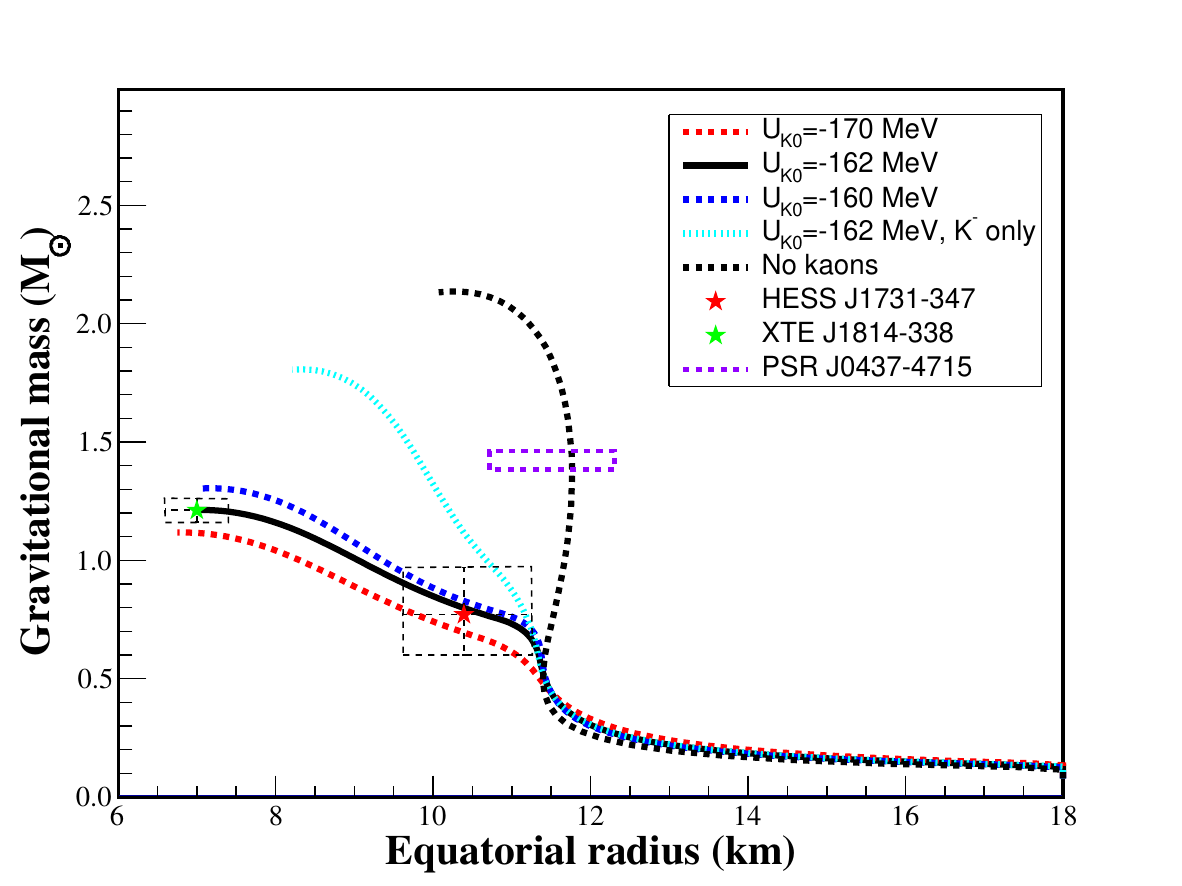}
\caption{Mass-Radius (M-R) diagram for several kaon potentials and without kaons. For the kaonic potential $U_{K0}$= - 162 MeV (both $K{^{-}}$ and ${\bar{K}^{0}}$ condensation) the dependence crosses directly across the central regions for the reported HESS J1731-347 (red star)\cite{HESS-2023} and the XTE J1814-338 (green star)\cite{Kini-2024} compact objects. The case considering only ${K^{-}}$ condensation is also depicted.}
\label{fgmr}
\end{figure}

Figure~\ref{fgeos} displays the dependence of the pressure on the energy density, where the onset of $K^{-}$ softens the EoS. Additionally, the onset of ${\bar{K}^{0}}$ condensate further softens the EoS at energy densities above 600 MeV/fm$^{3}$, corresponding to baryonic densities above 0.6 fm$^{-3}$, where both kaonic condensates already make a significant part of total mass density. 

This behavior is also depicted in Fig.~\ref{fig:pf_rmf}, where the particle fractions are illustrated as functions of the baryonic density. With the appearence of the $K^{-}$ onset on the $\beta$-equilibrium, the net $K^{-}$ field reduces the population of electrons and subsequently replacing them. Furthermore, when the conditions are favorable for the formation of ${\bar{K}^{0}}$, additional strangeness is introduced in dense matter without altering the charge neutrality as the fraction of ${\bar{K}^{0}}$ tracks the fraction of $K^{-}$ due to isospin symmetry.

Subsequently, Fig.~\ref{fgmr} denotes the gravitational mass as a function of radius (M-R diagram), for various values of kaon’s potential. 
The nucleonic branch reproduces both the astrophysical constraints for the maximum mass and the properties of the recently reported PSR J0437-4715 pulsar~\cite{PSRJ0437-4715}. 
It should be noted that the nucleonic branch shown in Fig.~\ref{fgmr} does not appear to describe the radius of the light compact object PSR J1231-1411~\cite{Salmi-2024} within the reported uncertainty ($R = 12.3-12.9$ km and $M = 1.04M_{\odot}$). However, due to convergence issues in the analysis and discrepancies when compared to other reported objects, this uncertainty may be underestimated.

In any case, a moderately redefined hadronic component on the EoS would achieve the reported properties while preserving the overall physical picture. 
When the onset of kaons emerges, the EoS is suitably soften, and as a result, reproduces the reported bulk properties of the light compact object XTE J1814-338. It is also noteworthy, that for $U_{K0}$ = -162 MeV, the EoS is able to reproduce both the XTE J1814-338 and HESS J1731-347 light compact objects, under the same theoretical framework. For completeness, the $K^{-}$ branch is also presented. Remarkably, the value 
$U_{K0}$ = -162 MeV is compatible with the value $U_{K0}$ = -180 $\pm$ 20 MeV, obtained in the study of kaonic atoms \cite{Friedman}. The presence of kaonic condensates in light compact objects can be considered as a relatively simple possibility to explain properties of such objects, without introducing more exotic scenarios such as hybrid stars, first-order phase transition in nuclear matter, or stars with admixture of dark matter.
The recent observation of the PSR J1231-1411 along
with the XTE J1814-338 object, where their masses are
comparable but the difference in radius is large ($\rm \sim 3.5~km$), enhances the argument that a hadronic EoS, which consists of a single nucleonic branch, is impossible to address these two objects. To this end, the role of the kaonic condensations can be important.  

The outcome conclusion of this paper is that a hadronic EoS, cannot account for two compact objects with similar masses but significantly different radii. To explain both objects simultaneously, it is essential to consider two distinct branches, each corresponding to a different composition of nuclear matter. The scenario concerning the possible existence of two branches in NSs is not new. First discussed in a work done by P. Haensel, M. Bejger and J. L. Zdunik~\cite{Haensel_2007}, where their goal was to investigate the Bethe \& Brown model prediction of $M_{NS}^{\rm max} = 1.5M_{\odot}$~\cite{Bethe-Brown_1994, Bethe-Brown_1995}, with the $2.1M_{\odot}$ NS reported by Nice et al.~\cite{Nice_2005} by introducing the two brances of NSs in order to reconcile a $2M_{\odot}$ pulsar and the SN 1987A. Within the same spirit later on, an extensive report concerning kaon condensation and its astrophysical implications was written by Gerald E. Brown, Chang-Hwan Lee and Mannque Rho~\cite{Brown-Rho_2008}.

In this context, we propose a possible explanation involving the onset of kaon condensation at high densities, which introduces a corresponding modification in the EoS. This adjustment could account for the existence of objects with relatively small masses while simultaneously exhibiting small radii.

\vspace{0.2cm} 
\noindent{\it{Acknowledgments.}} We would like to thank the reviewer for the fruitful and constructive suggestions that optimized our work. This work is supported by the Croatian Science Foundation under Project No. HRZZ-MOBDOL-12-2023-6026 and under the project "Relativistic Nuclear Many-Body Theory in the Multimessenger Observation Era" (No. HRZZ-IP-2022-10-7773), and by the Czech Science Foundation (GACR Contract No. 21-24281S). 

\bibliographystyle{elsarticle-num}

\end{document}